\documentclass[proof]{WileyASNA-v1}
\usepackage{natbib}

\articletype{Article Type}%

\received{20 September 2020}
\revised{.. .............. 2020}
\accepted{.. .............. 2020}

\begin{document}

\title{Near-horizon structure of escape zones of electrically charged particles around weakly magnetized rotating black hole: case of oblique magnetosphere}

\author[1]{Vladim\'{\i}r Karas*}

\author[1,2]{Ond\v{r}ej Kop\'a\v{c}ek}




\address[1]{\orgdiv{Astronomical Institute}, \orgname{Czech Academy of Sciences}, \orgaddress{\state{Prague}, \country{Czech Republic}}}

\address[2]{\orgdiv{Faculty of Science, Humanities and Education}, \orgname{Technical University of Liberec}, \orgaddress{\state{Liberec}, \country{Czech Republic}}}



\corres{*Bo\v{c}n\'{\i} II 1401, CZ-14100 Prague, Czech Republic. \email{vladimir.karas@cuni.cz}}


\abstract{We study the effects of large scale magnetic fields on the dynamics of charged particles near a rotating black hole. We consider a scenario in which the initially neutral particles on geodesic orbits in the equatorial plane become ionized, and hence they are destabilized by the charging process. Fraction of charged particles are then accelerated out of the equatorial plane and then follow jet like trajectories with relativistic velocities. We explore non axisymmetric systems in which the magnetic field is inclined with respect to the black hole spin. We study the system numerically in order to locate the zones of escaping trajectories and compute the terminal escape velocity. By breaking the axial symmetry we notice increasing fraction of unbound orbits which allow for acceleration to ultrarelativistic velocities.}




\jnlcitation{\cname{%
\author{Karas, V., et al.}} (\cyear{2021}), 
\ctitle{Near-horizon structure of escape zones of electrically charged particles}, \cjournal{Astronomische Nachrichten}, \cvol{XX; XX}.}

\maketitle


\section{Introduction}\label{sec1}
Nuclei of many galaxies are thought to harbour supermassive black holes. Strong gravity is manifested by a variety of effects, ranging from from very rapid motion of stars in dense nuclear clusters, shape of broadened skewed spectral lines from gaseous accretion disks, and other indirect evidence \citep{2016LNP...905.....H,2016ASSL..440.....B}. Furthermore, electromagnetic field plays an important role in shaping the gaseous structures and accelerating particles in the immediate vicinity of black holes in active galaxies \citep{2012bhae.book.....M,2015gimf.book.....K}. The system can be described by a set of mutually coupled Einstein-Maxwell equations within the framework of General Relativity. 

Although the presence of supermassive, strongly gravitating, dark compact objects is indicated by numerous independent approaches, the individual pieces of evidence for the black hole event horizon are challenging and not unique; new progress is expected with the help of  multi-messenger observations in the near future \citep[see][and further references cited therein]{2019NatRP...1..585M}. 

Mathematically rigorous and astrophysically relevant solution is defined with great precision by Kerr metric \citep{1963PhRvL..11..237K,Carter:1971zc}, where only the mass and angular momentum are free parameters that characterize the black hole and need to be measured by observation. Cosmic black holes are surrounded by gaseous environment and embedded in magnetic fields of external origin which are both essential for the process of mass accretion and light emission, however, these two ingredients have very little impact on resulting form of the gravitational field unless the black hole forms a compact binary system, in which case the dynamical space-time emerges and gravitational waves are released -- we do not consider the latter possibility in the present paper.

Because of high conductivity of plasma that forms accretion flows and due to strong differential rotation near the black hole, the effect of even weak electromagnetic fields is crucial. Near the equatorial plane, within the accretion torus, magnetic fields are turbulent and entangled on small length-scales $\ell$ (less than the geometrical thickness of the disk, $\ell\lesssim h(R)$); this leads to effective viscosity and drives the accretion process \citep{1991ApJ...376..214B}. On the other hand, in relatively empty funnels near the rotation axis, magnetic lines become organized on large scales (exceeding gravitational radius, $\ell\gtrsim R_{\rm g}$); this accelerates some plasma in collimated jets away from the black hole, so that only a small fraction of accreted material plunges into the event horizon \citep{2011MNRAS.418L..79T}. However, details of the mechanism responsible for this acceleration and collimation still remain a matter of debate \citep{2013rehy.book.....R}.

In \citet{2018ApJ...853...53K} we studied the role of a large-scale magnetic field aligned with the rotation axis of a rotating black hole. We assumed that an equatorial accretion disk orbits the black hole down to the innermost stable circular orbit \citep[ISCO, originally called the marginally stable orbit by][]{1972ApJ...178..347B} and we explored the proces of acceleration of electrically charged particles from the disk plane. In particular, we were interested in the terminal velocity that particles can reach under astrophysically realistic intensity of the aligned magnetic field (a few $\times 10^2$ gauss), and we found that the magnetic acceleration does operate, however, the particles typically reach only moderate Lorentz factors, $\gamma\lesssim3$. We also found that the regular (non-chaotic, integrable) motion typical for particles near an unperturbed (vacuum) black hole is broken; the imposed component of the external magnetic field leads to the emergence of zones of chaotic motion. Furthermore, in \citet{2020ApJ...900..119K} we relaxed the assumption of axial symmetry of the magnetic field and we studied the acceleration and chaoticity of motion by the magnetic field inclined with respect to the rotation axis. Interestingly, the oblique component leads to a more efficient acceleration and a larger terminal velocity, and a more complex (fractal) structure of the zones of chaotic motion. Let us emphasize that a combination of magnetic acceleration with the frame dragging by the black hole angular momentum are important. This happens because both the motion of particles as well as the shape of the magnetic field lines are affected by rotation.

In the present contribution we further explore how a moderately inclined magnetic field influences the terminal acceleration as a function of the black hole spin parameter ($|a|\leq 1$). We concentrate our attention to the case of co-rotation, $a>0$, and we focus on small radii, where the General Relativity effects operate most prominently. An interesting aspect of the adopted scenario is the fact that the acceleration process is most efficient in the vicinity of the plunging region boundary. It can thus be expected that the ejected particles form a hollow structure, which only further out spreads into a uniform collimated jet or an outflow. To this end the chaotic character of the outgoing trajectories enhances the mixing.

\section{Ejection of charged particles from magnetized ergosphere}

The spacetime metric coefficients of the Kerr black hole can be written in the well-known Boyer-Lindquist coordinate system \citep{Misner:1974qy,Chandrasekhar1983,Wald:1984rg}. The metric obeys the axial symmetry about the rotation axis and stationarity with respect to time, with a singularity hidden below the event horizon, where the curvature rises above all limits. Far from the event horizon, at spatial infinity the spacetime becomes asymptotically flat. All its mass $M$ is thus concentrated in the origin. The only other free parameter $a$ of the Kerr metric describes its rotation; the condition about the presence of the outer event horizon at a certain radius, $r=R_+$ (where the horizon encompasses the singularity) leads to the condition on maximum value of the dimensionless spin rate: $|a\leq1|$. However, the case of excessive spin and the naked singularity is not \textit{a priori} forbidden and the hereby described mathematical description remains in most of its aspects still valid without any change. The Kerr black hole solution can be then written in the form of the metric element \citep{Misner:1974qy,Chandrasekhar1983}
\begin{equation}
{\rm d}s^{2} = 
  -\frac{\Delta\Sigma}{A}\,{\rm d}t^{2}
   +\frac{\Sigma}{\Delta}\,{\rm d}r^{2}
   +\Sigma\,{\rm d}\theta^{2}
+\frac{A\sin^2\theta}{\Sigma}\;
   \left({\rm d}\phi-\omega\,{\rm d}t\right)^{2},
\label{metric}
\end{equation}
where $\Delta(r)=r^{2}-2r+a^{2}$,
$\Sigma(r,\theta)=r^{2}+a^{2}\cos^2\theta$, $A(r,\theta)=(r^{2}+a^{2})^{2}-{\Delta}a^{2}\sin^2\theta$, $\omega(r,\theta)=2ar/A(r,\theta)$ (geometrical units are assumed with the speed of light $c$ and gravitational constant $G$ set to unity). The above-given metric describes the vacuum spacetime, where the right-hand-side of the Einstein equations vanishes (no terms contribute to the energy-momentum tensor). This is clearly an astrophysical unrealistic assumption, nevertheless, it can be substantiated to certain precision if the amount of accreted matter is relatively small and the external electromagnetic fields are weak. Both assumptions are usually imposed, although this needs to be checked in each particular case. For example, the presence of a dense nuclear star-cluster and/or a massive accretion torus can change the gravitational field of the central black hole significantly; moreover the infall and a merger of a secondary black hole must lead to a non-stationary situation where gravitational waves are produced and the line element of the gravitational field is very different from the Kerr metric, albeit for only a limited period of time. 

Finding self-consistent solutions of mutually coupled Einstein-Maxwell fields is a difficult task because inherent non-linearity of the problem. Only a very limited class of exact solutions have been found under the simplifying assumptions about axial symmetry and stationarity \citep{1980esea.book.....K,1991JMP....32..714K,2000PhyS...61..253K}; otherwise one has to resort to numerical approaches. Fortunately, astrophysically realistic electromagnetic fields are usually weak with regard to their gravitational influence. One can thus explore test solutions of the Maxwell fields on the curved background of Kerr metric. The imposed spacetime is kept fixed, so that it does not evolve in time. Even if non-axisymmetric configurations cannot be stationary, the expected time-scale of the spacetime evolution are very long. On the technical side, this assumption means that we neglect all coupling terms of higher than the second order in electric and magnetic components.


On the black hole background the electromagnetic fields must be generated by currents flowing in the cosmic plasma far outside the event horizon. Within the limited volume around the black hole, the prevailing term corresponds to an asymptotically uniform magnetic field, which we will adopt hereafter. Even in this case, near the horizon the field line structure becomes increasingly entangled by the frame dragging \cite{1975PhRvD..12.3037K,1976GReGr...7..959B}. This structure becomes even more complex once the assumption about an axial symmetry is abandoned; the twisted field lines develop magnetic null points where reconnection events can occur \citep[see, e.g.,][]{2000EJPh...21..303D,2012CQGra..29c5010K}. In the adopted weak electromagnetic field limit the immediate consequence of the linearization is the fact that resulting electromagnetic four-vector can be written as a superposition of two parts $A^\mu=A_0^\mu+A_1^\mu$, the first one corresponding to an asymptotically uniform field aligned with the black hole rotation axis, and the other one corresponding to an asymptotically perpendicular configuration. Let us note that a small contribution to the black hole intrinsic electromagnetic field can originate from Kerr-Newman electric charge; however, this component is negligible due to rapid discharge by selective accretion of charged particles.

The case of uniform test magnetic field (corresponding to the aligned orientation) was first studied by \citet{Wald:1974np}. The magnetic field along the black hole rotation is described by two non-vanishing components of the four-potential,
\begin{eqnarray}
A_{0,t} &=& B_{0}a\Big[r\Sigma^{-1}\left(1+\mu^2\right) -1\Big], \label{mf1}\\
A_{0,\phi} &=& B_{0}\Big[{\textstyle\frac{1}{2}}\big(r^2+a^2\big)
 -a^2r\Sigma^{-1}\big(1+\mu^2\big)\Big] \sigma^2 \label{mf2},
\end{eqnarray}
in dimension-less spheroidal coordinates ($\mu=\cos\theta$, $\sigma=\sin\theta$). Eqs.\ (\ref{mf1})--(\ref{mf2}) represent an asymptotically homogeneous magnetic field.
The component perpendicular to the black hole axis has been given by \citet{1985MNRAS.212..899B,2007IAUS..238..139B},
\begin{eqnarray}
A_{1,t} &=& B_{1}a\Sigma^{-1}\Psi\sigma\mu, \label{mf3} \\
A_{1,r} &=& -B_{1}(r-1)\sigma\mu\sin\psi, \\
A_{1,\theta} &=& -B_{1}\Big[\big(r\sigma^2+\mu^2\big) a\cos\psi \nonumber \\
 && + \Big(r^2\mu^2+\big(a^2-r\big)(\mu^2-\sigma^2)\Big) \sin\psi\Big], \\
A_{1,\phi} &=& -B_{1}\Big[\Delta\cos\psi+\big(r^2+a^2\big)
 \Sigma^{-1}\Psi\Big] \sigma\mu, \label{mf4}
\end{eqnarray}
where $\psi\equiv\phi+a\delta^{-1}\ln\left[\left(r-R_+\right)/\left(r-R_-\right)\right]$, $\Psi=r\cos\psi-a\sin\psi$, $\delta=R_+-R_-$, and $R_{\pm}=1\pm\sqrt{1-a^2}$.
An arbitrarily inclined magnetic field can be obtained as superposition of the two above-given components. Their mutual relation defines the asymptotic angle of the magnetic lines of force. The set of four-potential vector components defines the structure of the electromagnetic tensor, $F_{\mu\nu}\equiv A_{[\mu,\nu]}$.

Let us note that the adopted solution for the black hole gravitational and electromagnetic fields is a very special one. Regarding the space-time metric, two Killing symmetries are imposed (stationarity and axial symmetry) and no further material is allowed to contribute to gravity (electro-vacuum test solution). These assumptions immediately exclude any significant mass in stars and the surrounding gas, and it also ignores the curvature arising from distant cosmological terms (an asymptotically flat metric is employed). Also the electromagnetic part of the solution represents merely the first term in the multipole expansion of the general solution for weak electric and magnetic intensities, which are stationary and organized on length-scales exceeding the gravitational radius (no time dependent, turbulent currents are allowed). However, we have relaxed the assumption about axial symmetry: the magnetic field at spatial infinity is inclined at an arbitrary angle with respect to the black hole rotation axis. 


\begin{figure}[tbh!]
\center
\includegraphics[scale=.29]{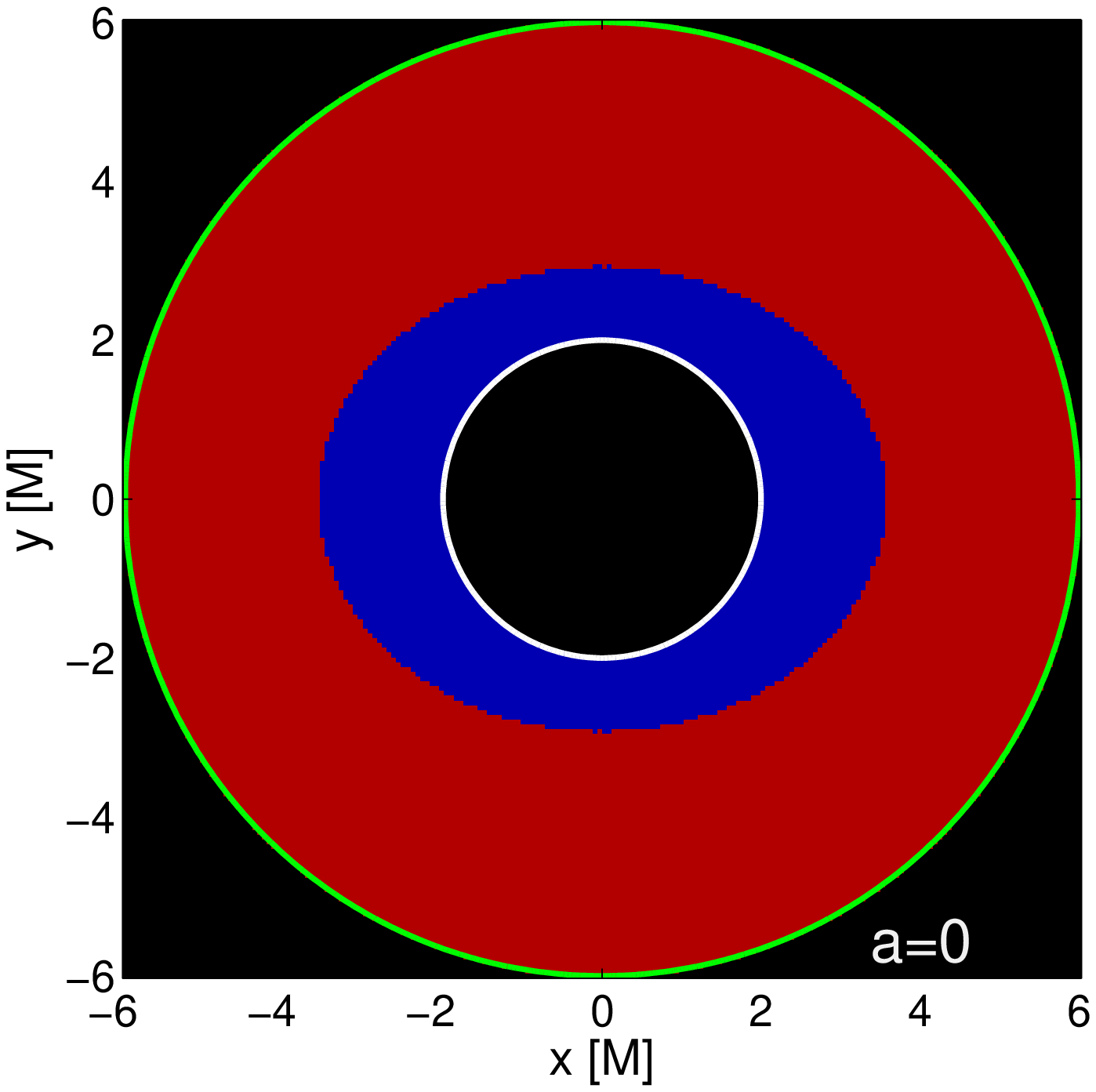}
\includegraphics[scale=.29]{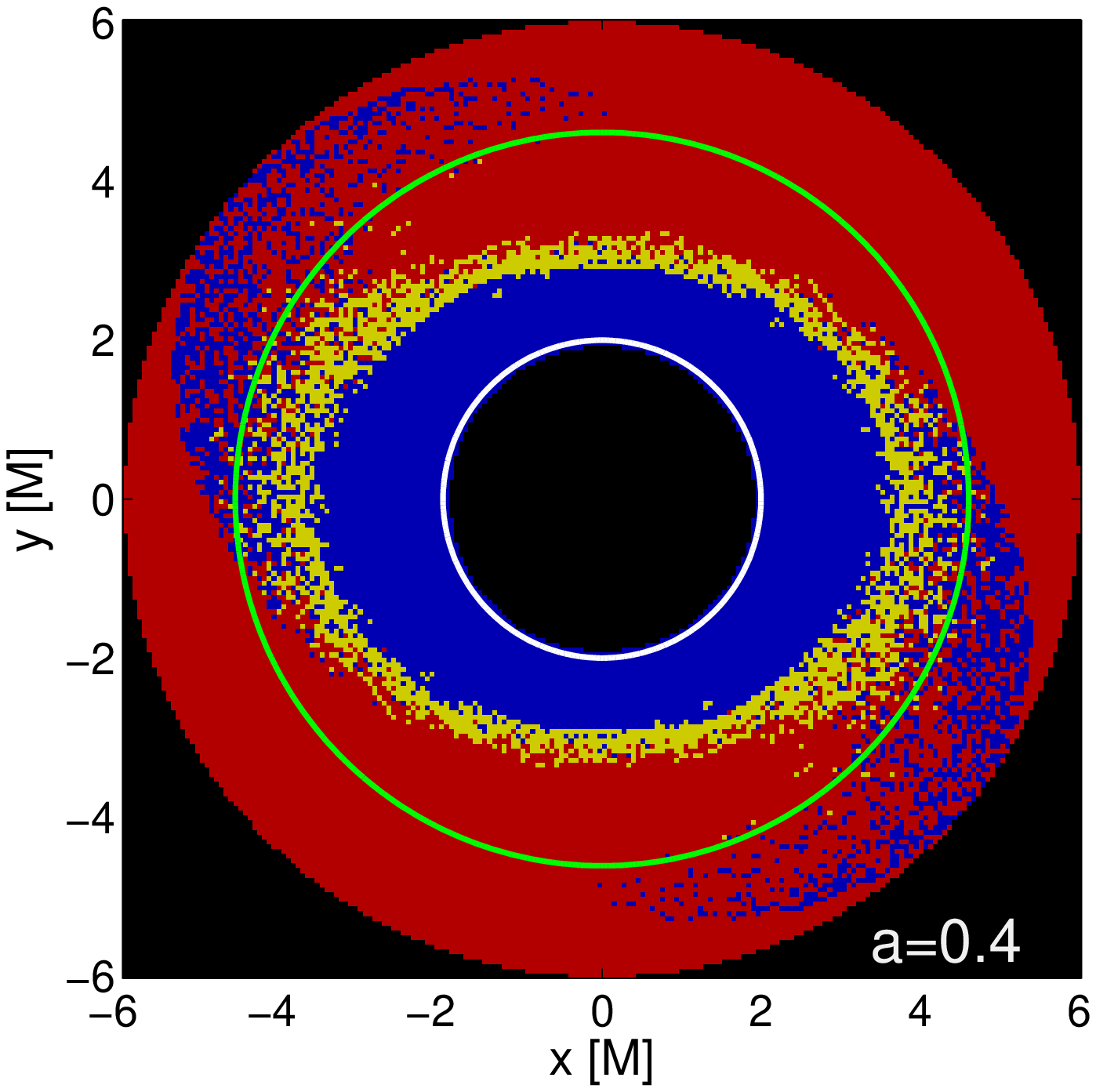}\\
\includegraphics[scale=.29]{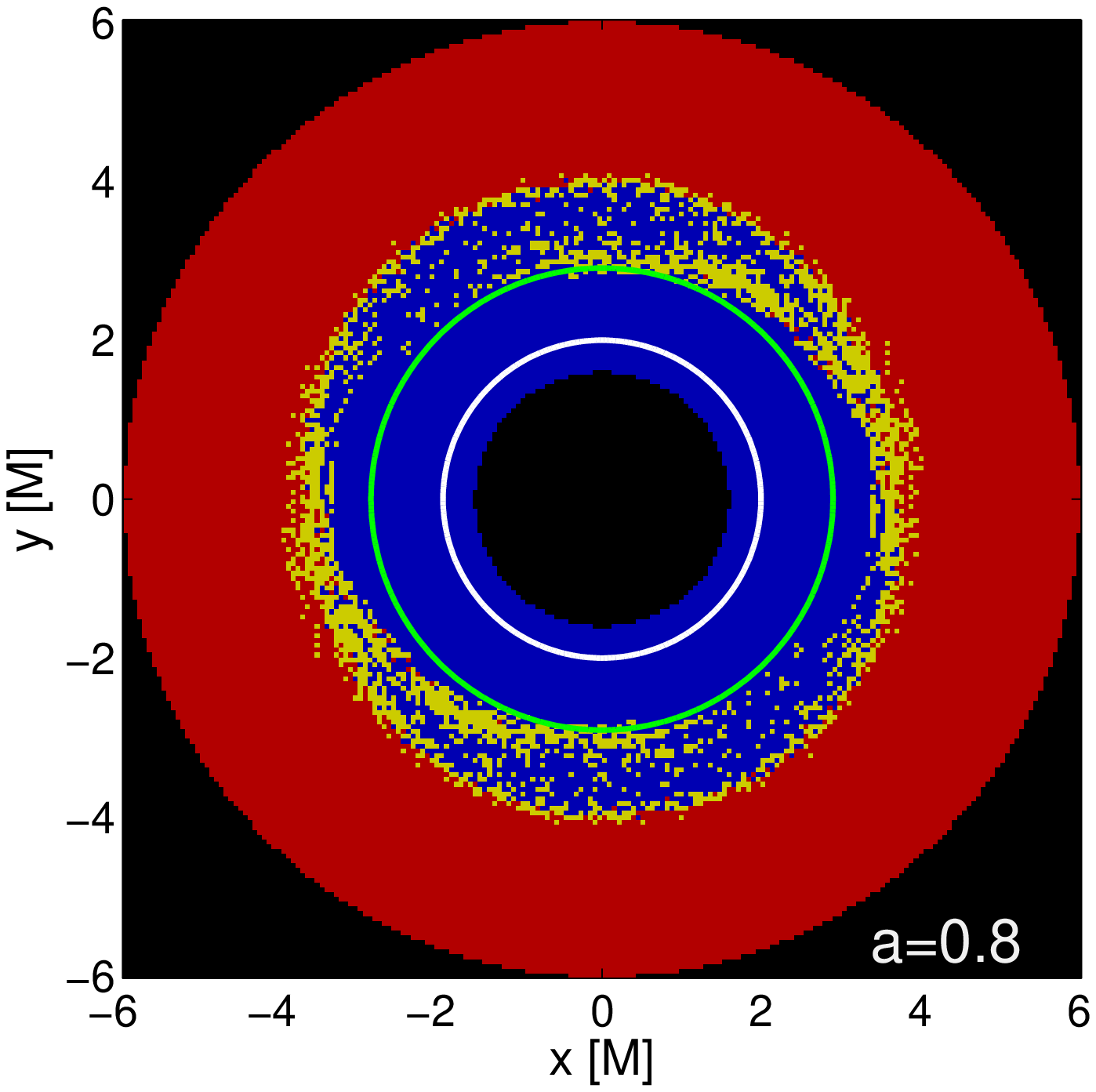}
\includegraphics[scale=.29]{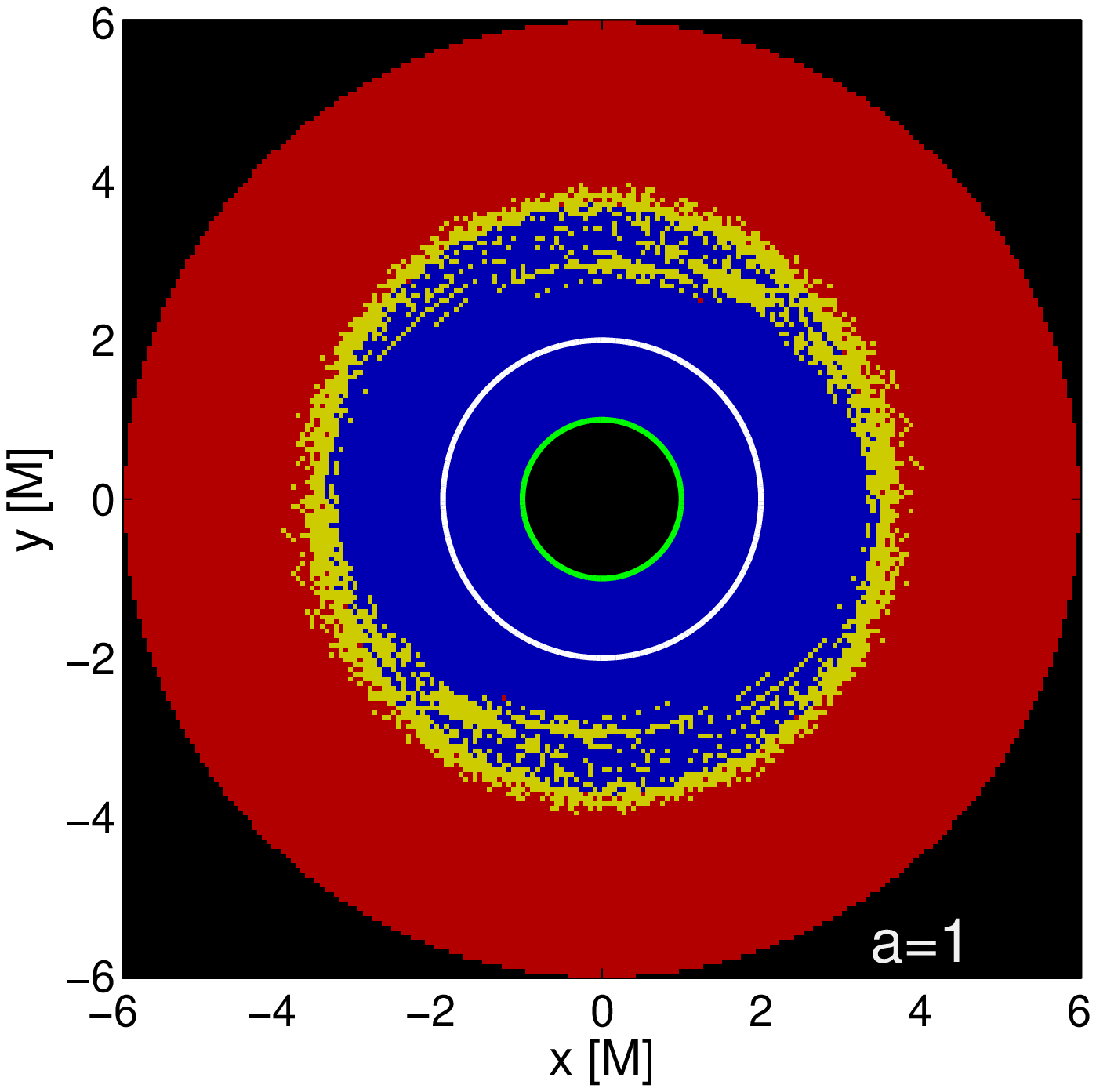}
\caption{The three classes of the orbits parameterized by the black hole spin $a$ and projected on the top view of the equatorial plane around the black hole. A hollow structure of escaping particles is visible (yellow). Color-coding: blue for plunging orbits, red for stable orbits, and yellow for escaping trajectories. The green circle denotes the innermost stable circular orbit (ISCO), which moves from $R_{\rm ISCO}(a=0)=6$ down to $R_{\rm ISCO}(a=1)=1$. The white circle is the boundary of the ergosphere. The inner black region marks the horizon of the black hole, $r=R_+(a)$. The other parameters of the system are $\alpha=35^{\rm o}$ and $qB=-5$. The asymptotic direction of the magnetic field is inclined in the positive $x$-axis direction.}
\label{fig1}
\end{figure}

The present paper builds on \citet{2020ApJ...900..119K}; we proceed systematically over the range of spin values and we demonstrate how the chaos gradually emerges in the particle motion. The transition to non-integrability is an interesting effect which requires the electromagnetic acceleration; it does not occur in the case of geodesic motion which is known to be completely integrable in Kerr metric \citep{2014ApJ...787..117K}.

\section{Motion within escape zones and the transition to chaos}

As mentioned above, we extend the numerous studies of the particle motion near black holes by considering one additional aspect that influences the trajectories in a distinct manner. This is the interplay between rotation of the black hole, as specified by the spin parameter $a$ (dimensionless angular momentum of the black hole $-1\leq a\leq1$), which defines the axis of symmetry of the gravitational field and the corresponding (perpendicular) equatorial plane, and an asymptotically uniform magnetic field, which is aligned along a different (arbitrary) angle of inclination $\alpha$ ($\alpha=0$ refers to the aligned configuration, whereas $\alpha=\pi/2$ the case of magnetic field perpendicular to the rotation axis). New features emerge which are not observed for particles near classical Kerr (rotating) and Kerr-Newman (rotating, electrically charged) black holes. Firstly, corridors of unbound escape motion occur for electrically charged particles that are injected into magnetospheres and start their acceleration on the fractal boundary region between plunging and bound trajectories. Secondly, these regions of chaotic motion develop where initially neighbouring trajectories separate exponentially from each other, as characterized covariantly in terms of properly defined Lyapunov coefficients and Poincar\'e sections \citep[and alternative signatures of chaos, such as the method of recurrence plots;][]{2010AIPC.1283..278K,2010ApJ...722.1240K}. 

Two main characteristics of the black hole spacetime are the mass $M$ and spin $a\equiv J/M^2c$, where J is the black hole angular momentum. Gravitational radius in physical units is then given by $R_{\rm g}\equiv GM/c^2\dot{=}1.5\times 10^5 M/M_\odot$ [cm], the outer horizon is located at radius $R_{\rm g}(1+\sqrt{1-a^2})$, and the ergosphere at $R_{\rm g}(1+\sqrt{1-a^2\cos^2\theta})$. Besides these canonical parameters, in our current model we have additional freedom in the above-mentioned magnetic angle $\alpha$, which relates two intensity components $B_0$, $B_1$, and the particle specific electric charge $q/m$. The system thus requires to specify a set of three independent parameters to be fully defined in geometrical units.

By introducing the external electromagnetic interaction the particle motion loses integrability. In consequence, we have to resort to numerical integration which needs to be performed systematically over the entire parameter space in order to reveal the emerging islands of chaos. For the purposes of the present investigation we start by setting the inclination angle $\alpha$ and magnetization parameter $qB$, and we integrate the system of equations of particle motion in their Hamiltonian form \citep{2020ApJ...900..119K},
\begin{equation}
\label{hamiltonian}
\mathcal{H}=\textstyle{\frac{1}{2}}g^{\mu\nu}(\pi_{\mu}-qA_{\mu})(\pi_{\nu}-qA_{\nu}),
\end{equation}
where $\pi_{\mu}$ is the canonical momentum and $g^{\mu\nu}$ contravariant components of the metric tensor. The equations of motion then read
\begin{equation}
\label{hameq}
\frac{{\rm d}x^{\mu}}{{\rm d}\lambda}\equiv p^{\mu}=
\frac{\partial \mathcal{H}}{\partial \pi_{\mu}},
\quad 
\frac{d\pi_{\mu}}{d\lambda}=-\frac{\partial\mathcal{H}}{\partial x^{\mu}},
\end{equation}
with $\lambda\equiv\tau/m$ being the affine parameter and $\tau$ proper time along the trajectory.

\begin{figure*}[tbh!]
\center
\includegraphics[scale=.24]{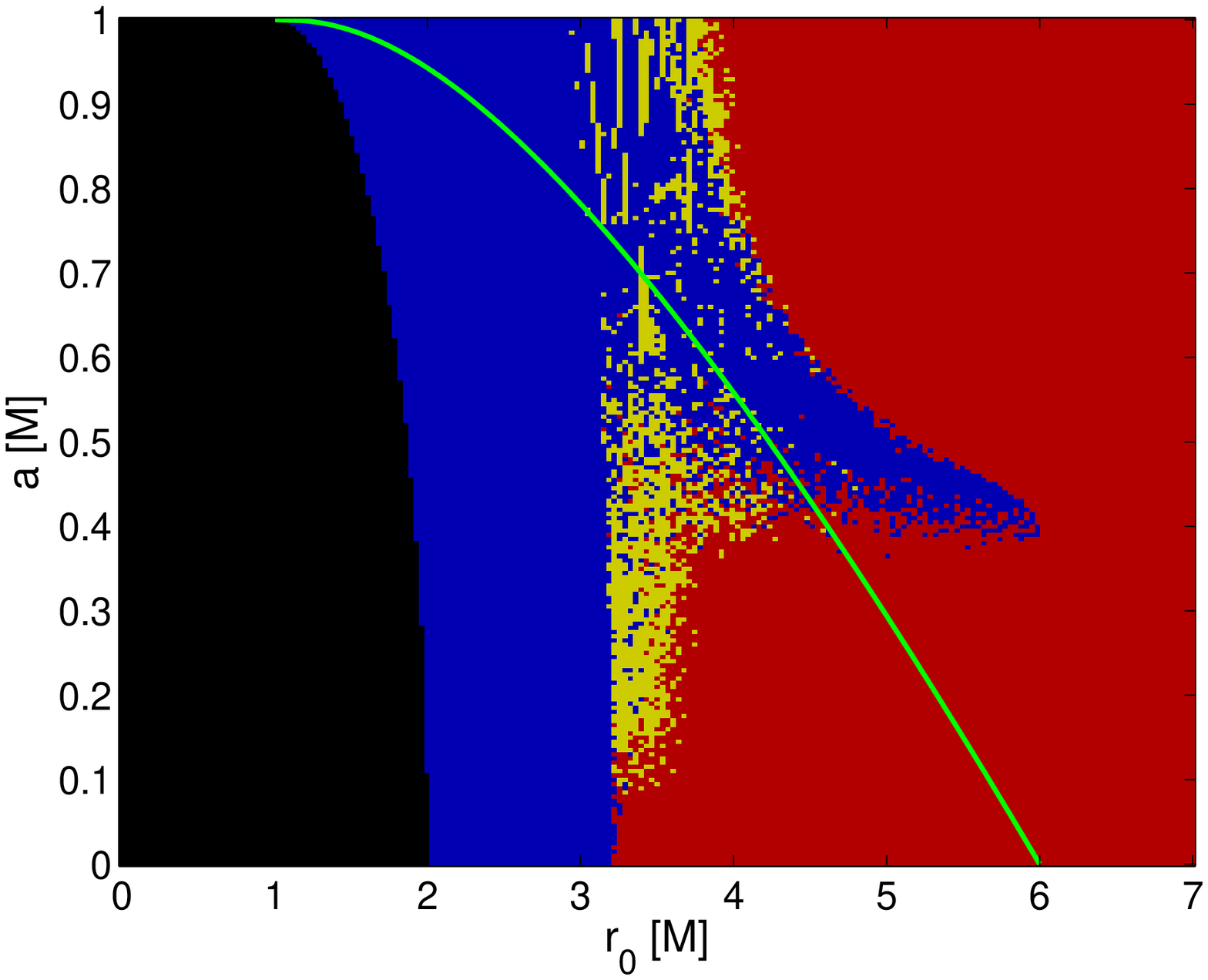}
\includegraphics[scale=.24]{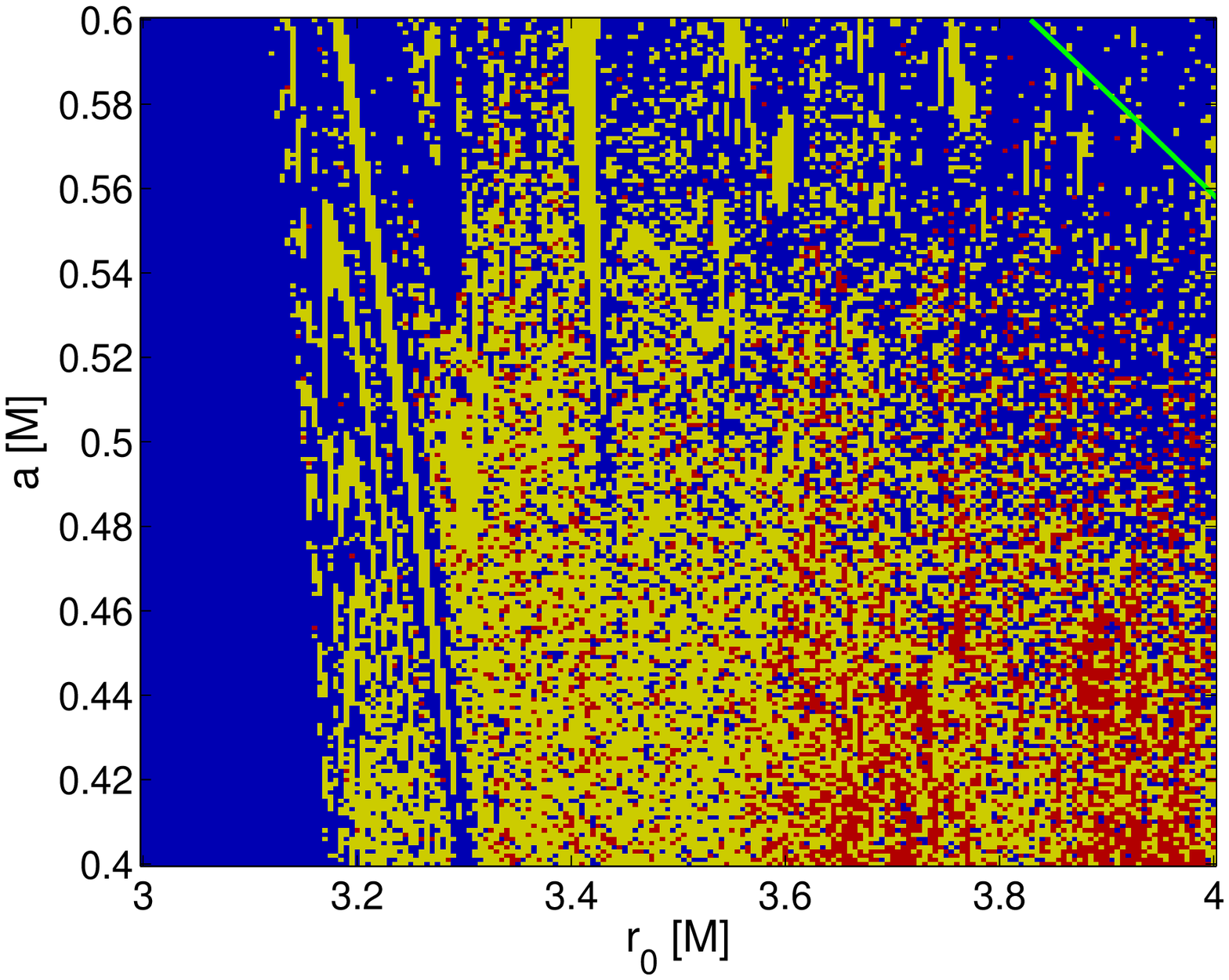}
\includegraphics[scale=.24]{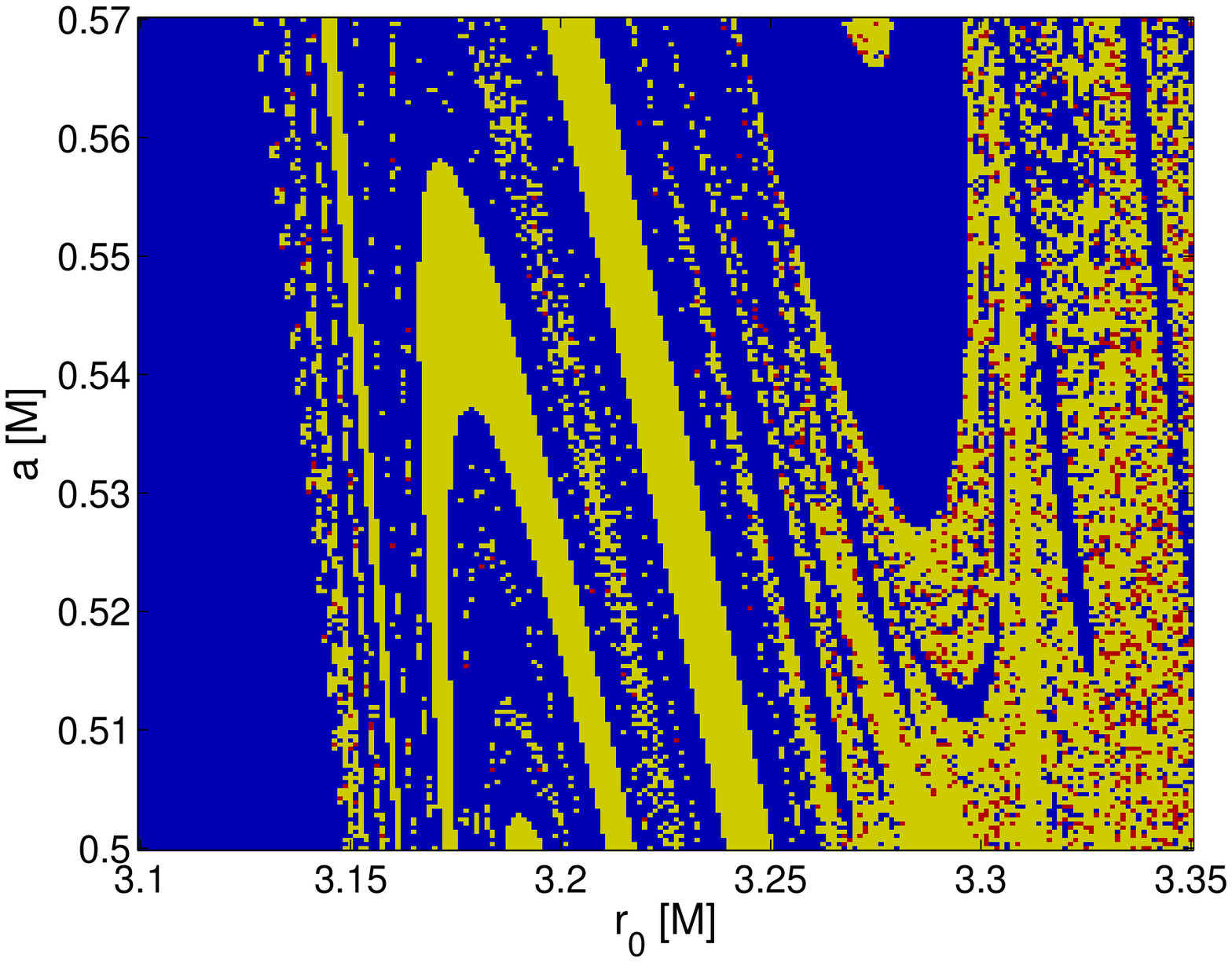}
\includegraphics[scale=.24]{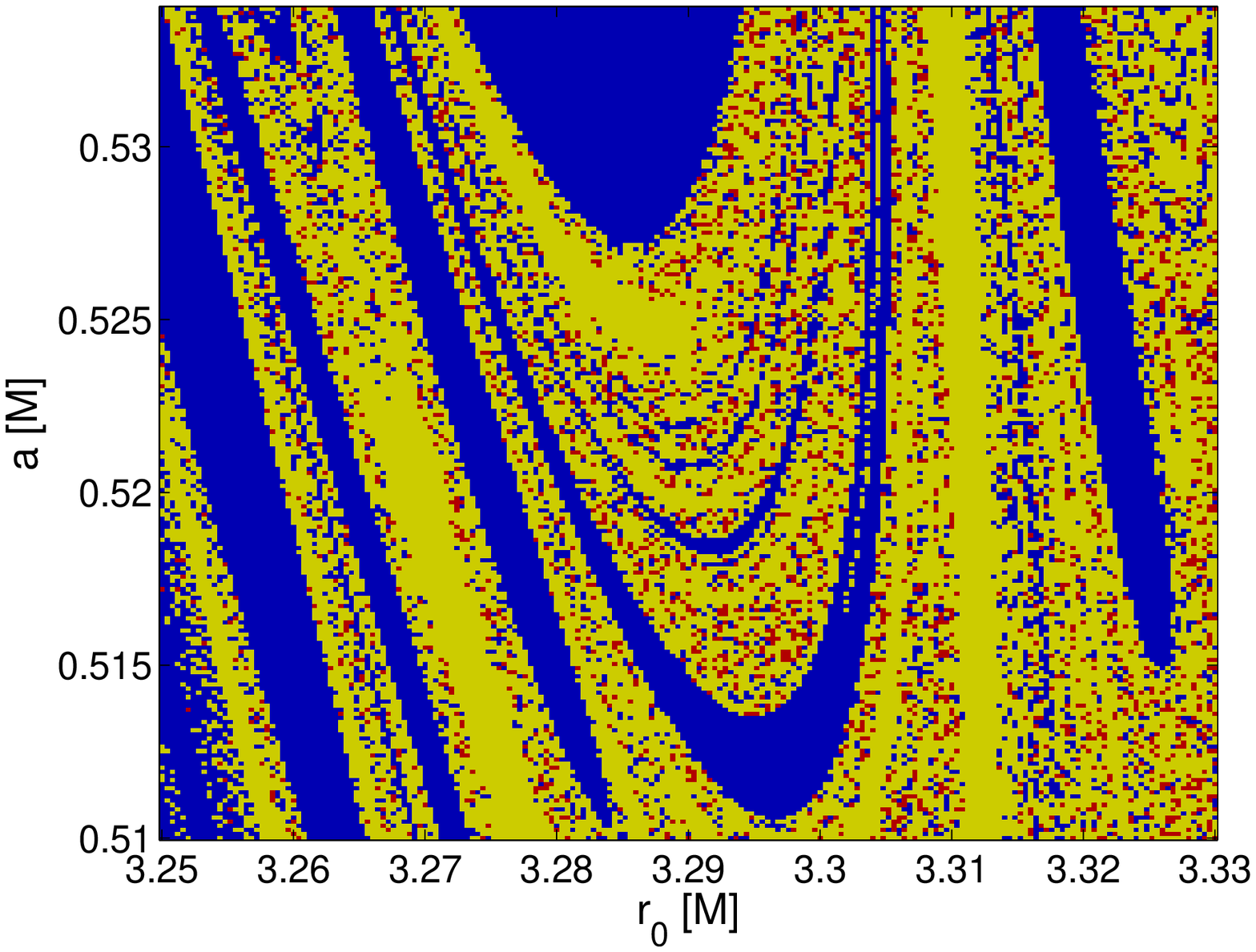}
\caption{The fractal structure emerges of escaping, plunging, and bound trajectories with the rising value of spin $a$. Gradually increasing resolution is shown going from left to right. Parameters: $\alpha=35^{\rm o}$, $qB=-5$ and $\varphi_0=135^{\rm o}$.}
\label{fig2}
\end{figure*}

Figure \ref{fig1} depicts three different types of trajectories that start from the black hole equatorial plane $(x,y)$ at some initial radius $r=r_0$. These trajectories are distinguished into three categories according to the final fate of the particle: (i)~plunging orbits which enter into the black hole horizon; (ii)~stable orbits bound to the black hole while avoiding accretion; and (iii)~ejected orbits along which particles can escape away from the black hole to radial infinity. Distinct features can be observed, which are obviously invariant with respect to the choice of coordinate system. Firstly, the escaping trajectories start to emerge for non-zero values of spin and they exist up to $a\rightarrow1$. These orbits spatially occur approximately in between stable and plunging trajectories; however, the image suggests an interesting fractal structure. This would be naturally expected on the basis of the fact that some selected trajectories in the escaping region appear to be chaotic by numerical integration.
 
In order to reveal details of this complex region we can zoom progressively into a small parts of the phase space. An example is shown in Figure \ref{fig2}$\!\!$, where we plot the three classes of the trajectories in $(a,r_0)$ plane. Let us note that the region of escaping orbits can come up above ISCO at certain value of $a\approx0.5$. However, for high values of $|qB|$ the resulting escape zone can even get inside the ergosphere, although the region is then very narrow. Because the plots have been generated by numerical integration, a natural question has to be raised regarding the precision and the sensitivity to initial conditions. We integrated the motion equations in the hamiltonian form via Adams-Bashforth-Moulton multi-step integrator, as described in a recent paper \citet{2020ApJ...900..119K} in more detail. The routine is based on the predictor-corrector loop with the adaptive step-size and the local truncation error controlled by a relative tolerance parameter. Further, to verify the precision, we also employed Dormand \& Prince explicit Runge-Kutta type scheme \citep[{\sf ode873;}][]{1978CeMec..18..223D,prince1981}. Comparing the output we conclude that the overall structure produced by the two independent schemes is consistent and the regions of different classes of trajectories agree. However, let us also note that caution always has to be paid with respect to the precision. For example, the standard fourth-order Runge-Kutta integration routines are not adequately accurate to reflect the regions of chaos \citep{2014ragt.conf..123K}.

\begin{figure*}[tbh!]
\center
\hfill~
\includegraphics[scale=.36,trim={20mm 0 15mm 0},clip]{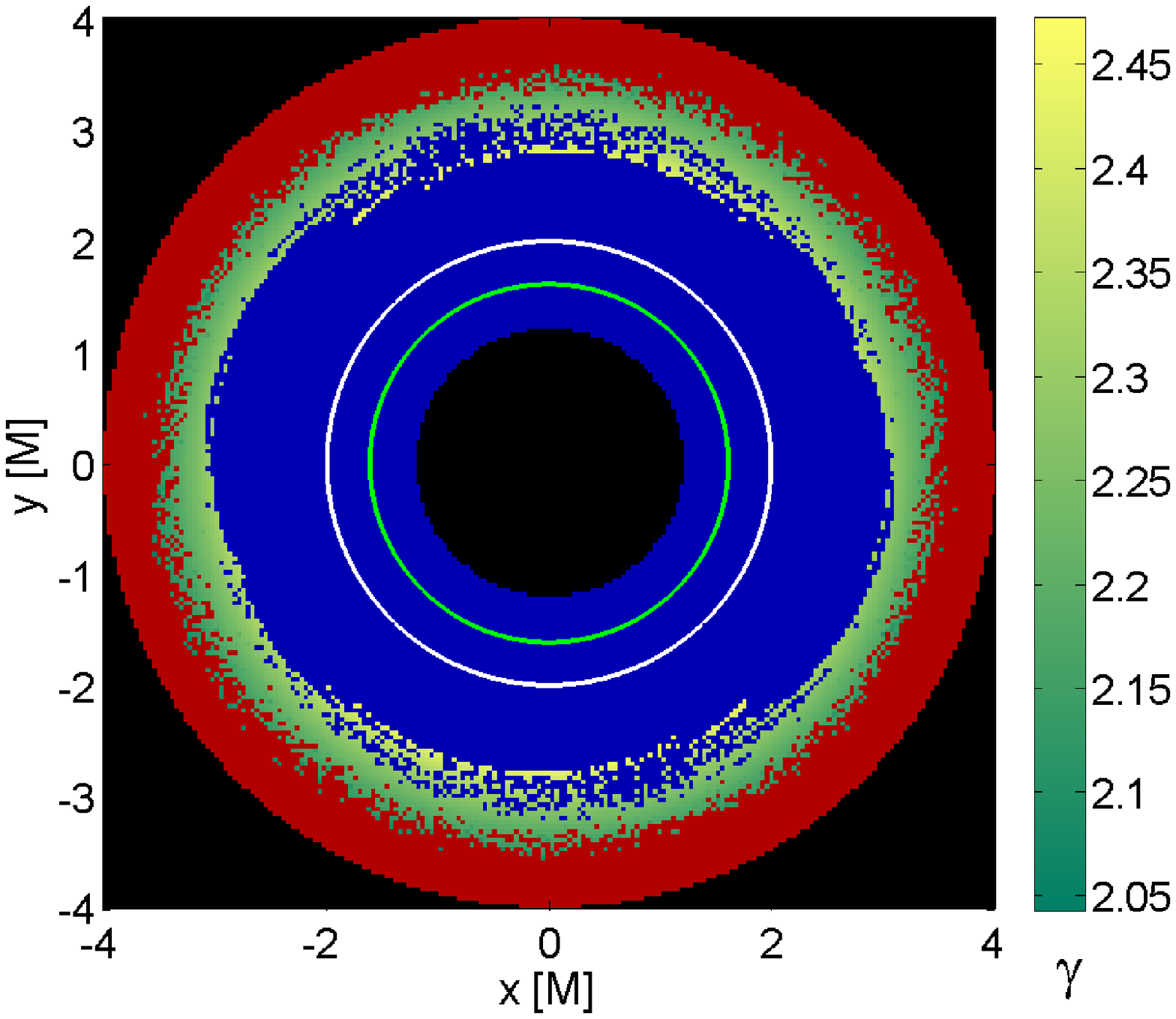}
\hfill~
\includegraphics[scale=.36,trim={20mm 0 21mm 0},clip]{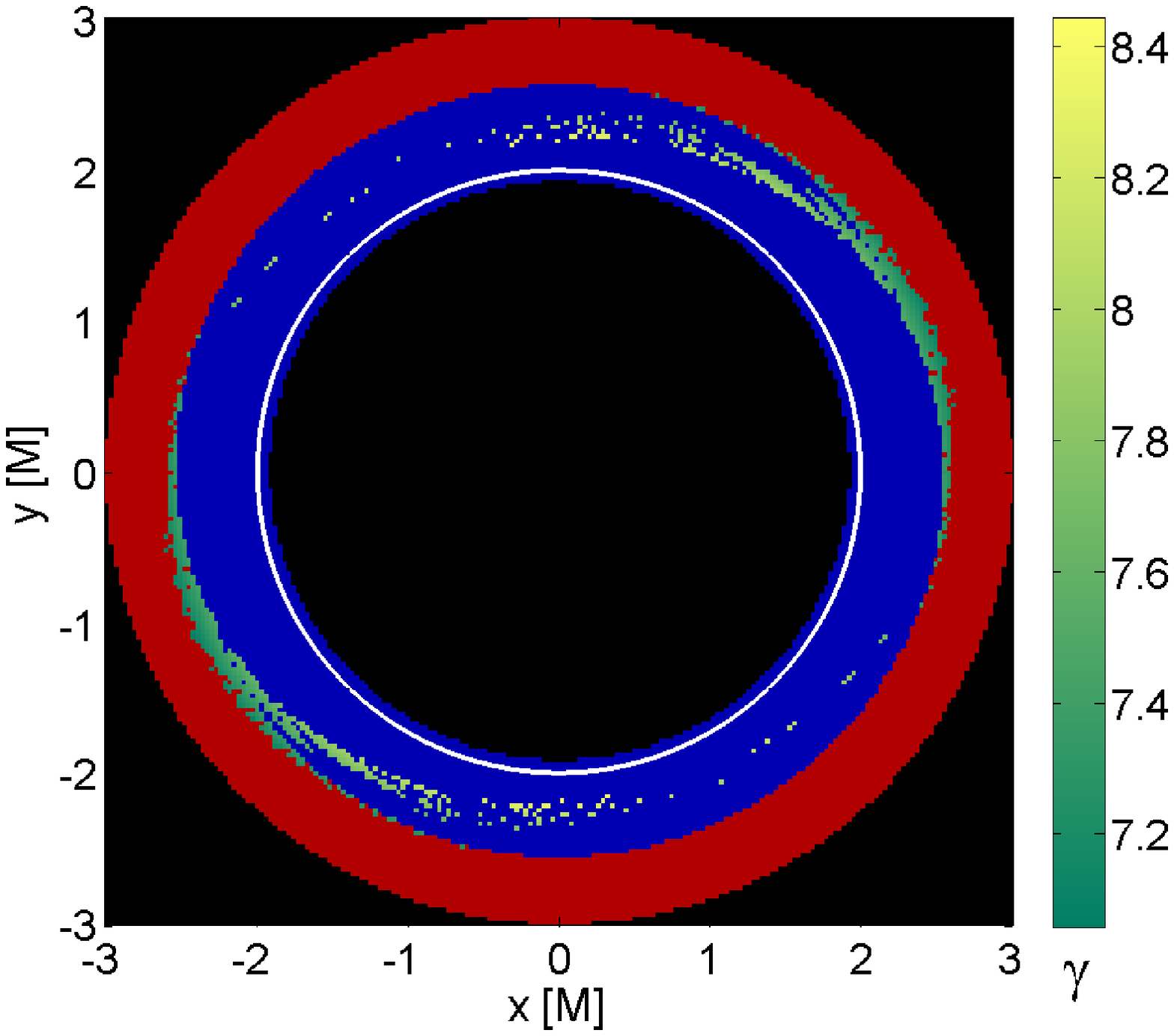}
\hfill~
\caption{Final Lorentz factor $\gamma$ of the escaping trajectories. Values of several parameters have been kept fixed. Left panel: moderate angle of the magnetic field $\alpha=25^{\rm o}$, charge and magnetization $qB=-5$, and almost maximally rotating Kerr hole $a=0.98$. Right panel: $\alpha=35^{\rm o}$, $qB=-50$, $a=0.4$.}
\label{fig3}
\end{figure*}

Regarding the class of escaping trajectories an interesting question concerns the final velocity that the ejected particles can reach. To this end we construct a colour-coded map of the terminal Lorentz factor $\gamma$ in Fig. \ref{fig3}$\!\!$. In this particular example we set $qB_0<0$ (a necessary condition for the particle escape). The main ingredients that are necessary and sufficient to launch the outflow of charged particles are a combination of rotation of the central black hole with the imposed large-scale magnetic field. A moderately misaligned configuration increases the efficiency of acceleration. On the other hand, a perfect alignment (magnetic parallel with the spin, for the same values of all other parameters) restricts considerably the efficiency of the acceleration mechanisms. 

The escape zone is clearly distinguished in the plot by yellow to green colours, where the acceleration reaches up to $\gamma\approx 2.5$ for the given set of other (frozen) parameters. The plots confirm that the escaping particles originate from a rather narrow range of radii between $2.5\lesssim r\lesssim 3.5$. This is a generic picture which appears typically for moderate $\alpha\lesssim45^{\rm o}$. On the other hand, for $\alpha\gtrsim45^{\rm o}$ we found that the particles can escape from only certain azimuthal locations \citep[for further details, see also][]{2020ApJ...900..119K}.
 
 \section{Discussion and conclusions}
We further examined aspects of charged particle acceleration in the region of organized magnetic field near rotation axis of Kerr black hole. While the magnetic field develops a prevailing toroidal component in the equatorial torus, where they are highly turbulent, in the diluted above the torus the large/scale poloidal filed lines help to accelerate a stream of outflowing particles which eventually form a collimated jet. This picture has been suggested by several independent numerical simulations \citet[e.g.][]{2011ApJ...732L...6R}. It turns out that a moderate (non-zero) inclination of the large-scale magnetic field helps the acceleration; the effect operates less efficiently in the axially symmetric (aligned, $\alpha\rightarrow0$) configuration, and it is also less efficient in the highly inclined (or even perpendicular, $\alpha\rightarrow\pi/2$) case. We find that the ejection mechanism requires an interplay between rotation of the black hole and the magnetic field. The particles originate from the region in the equatorial plane near above the outer boundary of the ergosphere and below the innermost stable circular orbit, therefore very close to the horizon.

The ejection region exhibits fractal structure and the trajectories show signatures of chaos. Although we had to resort to numerical integration of the trajectories, we verified the precision of our method by employing two highly accurate schemes and checking the results independently by recurrence analysis; all approaches qualitatively agree. The emergence of chaos is an interesting feature in the black hole spacetime, where the orbits of the unperturbed Kerr metric are known to be fully integrable. Although any direct observation confirmation about the character of motion is difficult or impossible because of an insufficient resolution, it is interesting to recall the hollow structure of the wobbling jet in the core of M87 elliptical galaxy \citep{2016ApJ...817..131H,2017A&A...601A..52B}. Here, the resolution of the interferometric images reaches down to the horizon scale, and it indeed appears that the material of the jet might start from around the inner rim of the accretion torus, corresponding to the region indicated by the ejected (yellow) orbits in our graphs.
  
\section*{Acknowledgments}
We thank the \fundingAgency{Czech Science Foundation}, grant ref.\ \fundingNumber{19-01137J} and the \fundingAgency{Czech Ministry of Education, Youth and Sports} COST program ref.\ \fundingNumber{LTC\,18058} to support international collaboration in relativistic astrophysics.

\bibliography{Karas}

\end{document}